# Visualization for Periodic Population Movement between Distinct Localities

Alexander Haubold[*]
Department of Computer Science
Columbia University

## Abstract

We present a new visualization method to summarize and present periodic population movement between distinct locations, such as floors, buildings, cities, or the like. In the specific case of this paper, we have chosen to focus on student movement between college dormitories on the Columbia University campus. The visual information is presented to the information analyst in the form of an interactive geographical map, in which specific temporal periods as well as individual buildings can be singled out for detailed data exploration. The navigational interface has been designed to specifically meet a geographical setting.

## 1 Introduction

Visualization of large, highly dimensional data sets is an important and essential form of data analysis that applies to every field of information analysis. It lends itself especially well to mapping the temporal movement of people between distinct localities in a well-defined area, such as a university campus, a city, country, or the world. Generic visualizations of this type are widely used in historical [1] and socio-economic [2] contexts. In the context of this paper, Columbia University's Residence Hall (URH) administration sought a tool to visually evaluate the movement of students between dormitories on campus; they provided the needs and questions that informed the visual display we have developed. While the raw data is available to the administration, it has never been used for analytical purposes, because there exists no tool that quickly discerns the data for useful results.

Throughout the design we have tried to pick visual attributes to draw attention to the things analysts cared about the most. This breaks down into two distinct approaches: 1. purely visual techniques, and 2. metaphorical, mnemonic associations between images and their representation, creating a type of visual semantic.

## 2 Interface Design

The visualization interface features a two-dimensional interactive geographical map of city blocks and buildings that assume different states, as well as directed "relocation" arcs that represent relocations between two locations (Figure 1). In populating the map, we have paid careful attention to something we call a "contrast budget" as well as the order in which graphical components are placed on the map. A minimal portion of contrast has been set aside to manage the information provided by the system, while the larger portion is used to manage the viewer's input. We also use hue and saturation to distinguish different types of visual representations. City blocks and data-irrelevant buildings have been colored in a low contrast and close to monochrome value, as their role is to merely provide spatial context. Buildings associated to data are emphasized in a separate color. As buildings become more interesting to the analyst (as evidenced by their being selected, armed, or selected and armed) the saturation changes exponentially to reflect the attention the viewer has given to the object.

---
[*]e-mail: ah297@columbia.edu

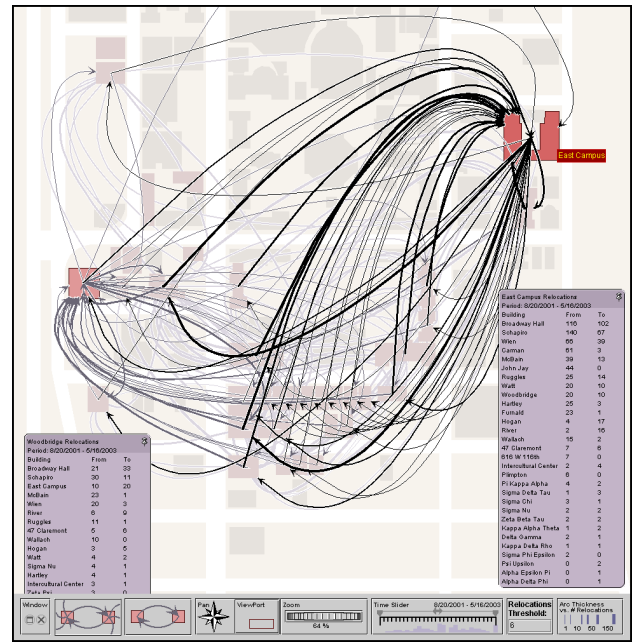

**Figure 1.** Interface. City blocks and data irrelevant buildings appear in background colors; data relevant buildings and relocation arcs each assume specific color value, while their saturation changes according to user input and user interest. Movable relocation summary cards present detailed numerical data for each selected building.

Relocation arcs follow a similar trend in increasing contrast versus increasing importance, and are additionally distinguished by their placement and mode of appearance. Links that are not associated to armed or selected buildings generally appear on the same background level as city blocks and irrelevant buildings. Furthermore, only links of substantial relocations are shown in the background, the threshold of which can be changed interactively. Arc thickness is adjusted logarithmically to the number of relocations in order to preserve the distinctiveness of the arcs. As buildings and their associated relocations become more interesting for the viewer, the arcs move into a position closer to the foreground, while at the same time assuming more saturated values.

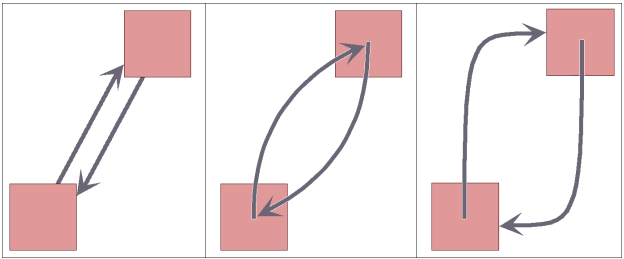

**Figure 2.** Relocation Links. Left: Straight lines; Middle: Symmetric arcs; Right: Spiral-shaped arcs homing in onto target object.

Relocations links between two given buildings appear in a clock-wise directed fashion while the spiral-shaped arcs sharply home in on the target object. The design of relocation arcs went through several iterations (Figure 2). First, simple straight lines were curved concavely to visually separate the relocation links from the inherently boxy building nodes. By means of this method we distinguish nodes from links as early as possible in visual processing. In a second step, we have changed the profile of the curve from a simple circular arc to one with an ever increasing curvature. This makes it easier to visually distinguish the beginning from the end of an arc, and complements the use of a directed arrow.

## 3 Interface Tools

A two-sided time slider, a more general version of which has been introduced by [3], features the distinct time periods over which relocation data exists, and allows the viewer to specify a lower and upper bound for displaying relocations during a particular period (Figure 3). The left, right, and middle arrows can be moved independently to increase the lower bound, increase the upper bound, and move a constant time period in either direction, respectively. Featured below the time line is a histogram of total relocations for each time period, where the values are adjusted logarithmically. Given the reduced space for the histogram, a linear scale would single out only the periods with high activity, which results in a too sparsely populated histogram.

For each selected building a relocation summary card appears in the interface, which gives a numerical summary of the relocation data over the selected time period. This card can be moved freely within the interface and pinned onto the map similar to a PostIt note. A similar Details on Demand method utilizing "info cards" has been first used in [4].

## 4 Data Model and Interoperability

The interface is not restricted to the specific data presented herein. In a one-time pre-process step (Figure 4), a bitmapped geographical map is automatically vectorized, resulting in a list of polygons with corresponding fill color values. A second text file enumerates each color and maps colors to building names. A third file enumerates relocation matrices for each time period.

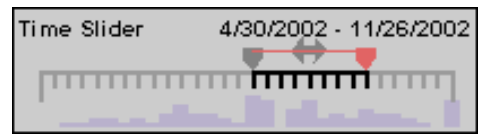

**Figure 3.** Time Slider with embedded histogram.

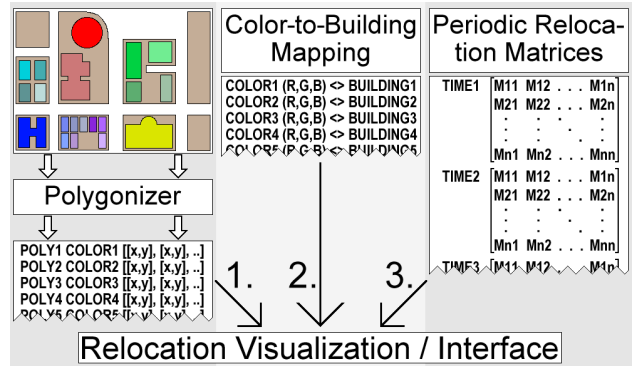

**Figure 4.** The Relocation Visualization is generated using a polygonized bitmap, a color-to-building map, and a periodic relocation matrix file.

These three text files serve as the input to the visualization interface. Using this data model, any geographical area can be presented in the visualization tool, including cities, building floor plans, and also material of non-geographical nature.

## 5 Conclusion

We have developed an information visualization method and a practical tool to aide in analyzing periodic movement between buildings (or other entities) within a defined spatial region. Using different conceptual layers, the information is presented to the viewers in a passive overview while giving them interactive tools to filter out buildings and associated relocations of interest. As this is a work in progress, we are further exploring which visual attributes are best suited for the purposes of visualization and interaction.

## Acknowledgements

Discussions with W. Bradford Paley were the source of the spiral arcs and color choices and were fruitful in helping to keep the visual representations driven by the needs and expectations of the analysts rather than just the structure of the data.